\documentclass[twocolumn,showpacs,preprintnumbers,amsmath,amssymb]{revtex4}

\usepackage{graphicx}
\usepackage{dcolumn}
\usepackage{bm}

\begin{document}

\title{Finite time decoherence could be suppressed efficiently in photonic crystal}

\author{Fa-Qiang Wang}
\author{Zhi-Ming Zhang}
\email{zmzhang@scnu.edu.cn}
\author{Rui-Sheng Liang}%
\affiliation{Lab of Photonic Information Technology, School of Information and Photoelectronic Science and Engineering, South China Normal University, Guangzhou 510006, China}%

\date{\today}

\begin{abstract}
The decoherence of two initially entangled qubits in anisotropic
band gap photonic crystal has been studied analytically without Born
or Markovian approximation. It is shown that the decoherence
dynamics of two qubits in photonic crystal is greatly different from
that of two qubits in vacuum or subjected to usual non-Markovian
reservoir. The results also show that the finite time decoherence
invoked by spontaneous emission could be suppressed efficiently and
the entanglement of the Bell state possesses odd parity is more
easily preserved in photonic crystal than that of the Bell state
possesses even parity under the same condition. A store scheme for
entangled particle pair is proposed.
\end{abstract}

\pacs{03.67.mn,03.65.Yz,03.65.Ud,42.70.Qs}
\maketitle
\par Entanglement is recognized as a global quantum-mechanical
effect and has played a key role in quantum information\cite{ab,sf},
quantum computation\cite{ab1,lk}, quantum cryptography\cite{ak}, and
so on. However, the phenomenon , termed as ``entanglement sudden
death"(ESD), has been found theoretically\cite{yu1,yu2} and shown
experimentally\cite{mp,mf}. It is shown that spontaneous
disentanglement may take only a finite-time to be completed, while
local decoherence (the normal single-atom transverse and
longitudinal decay) takes an infinite time\cite{yu2}. And some
issues have been devoted to extending the results in Markovian
regime to non-Markovian case\cite{cao,bel}.
\par On the other hand, photonic crystals form a new class of dielectric materials,
in which the electromagnetic interaction is controllably altered
over certain frequency\cite{john3}. The periodic dielectric
structures leads to the formation of a photonic band gap( PBG), a
range of frequencies for
 which no propagating electromagnetic modes are allowed \cite{john3}.
 The presence of the photonic band gap in the dispersion relation of
 the electromagnetic field results in a series of new phenomena, including the inhibition
 of the spontaneous emission \cite{ey1}, strong localization of light\cite{john4}, formation
 of atom-photon bound states \cite{john5}. In this Letter, we will
 investigate how the entanglement of two qubits evolute in photonic
 crystal.
\par Now we restrict our attention to two noninteracting two-level
atoms A and B coupled individually to two photonic crystal
environment reservoirs which are initially in vacuum states. To this
aim, we first consider the Hamiltonian of the subsystem of single
qubit coupled to its reservoir as
\begin{equation}
H=\omega_{0}|e\rangle\langle
e|+\sum_{k}\omega_{k}a_{k}^{\dagger}a_{k}+\sum_{k}g_{k}\left(a_{k}^{\dagger}|g\rangle\langle
e|+a_{k}|e\rangle\langle g|\right)
\end{equation}
where $\omega_{0}$ is the atomic transition frequency between the
ground state $|g\rangle$ and excited state $|e\rangle$. The index
$k$ labels the field modes of the reservoir with frequency
$\omega_{k}$, $a_{k}^{\dagger}$ and $a_{k}$ are the modes' creation
and annihilation operators, and $g_{k}$ is the frequency-dependent
coupling constant between the transition $e-g$ and the field mode
 $k$. For a single excitation of the subsystem, these states
are
\begin{eqnarray}
\psi_{_{1}}&=&|e\rangle \bigotimes \prod_{k}|0_{k}\rangle \nonumber \\
\psi_{_{k}}&=&|g\rangle \bigotimes |1_{k}\rangle\prod_{k^{'}\neq
k}|0_{k^{'}}\rangle \label{one}
\end{eqnarray}
where the ket $|0_{k}\rangle$ indicates the field mode $k$ is in
vacuum state. $|1_{k}\rangle$ indicates the field mode $k$ is in the
first excited state. The unexcited state
\begin{equation}{\label{two}}
\psi_{0}=|g\rangle|\bigotimes\prod_{k}|0_{k}\rangle
\end{equation}
is not coulped to any other state. \par Now we could expand a
general state vector of the subsystem as\cite{garr}
\begin{equation}
    \psi(t)=c_{0}\psi_{0}+c_{1}e^{-i\omega_{0}t}\psi_{1}+\sum_{k}c_{k}e^{-i\omega_{k}t}\psi_{k}
\end{equation}
in terms of the states (\ref {one})and (\ref {two}
)and insert this
into the schr\"{o}dinger equation $i(d/dt)\psi=H\psi$ to obtain the
following set of coupled equations:
\begin{eqnarray}
 \dot{ c_{1}} &=& -i\sum_{k}g_{k}e^{-i(\omega_{k}-\omega_{0})t}c_{k} \label{three}\\
  \dot{c_{k}} &=& -ig_{k}e^{i(\omega_{k}-\omega_{0})t}c_{1}\label{four}
\end{eqnarray}
The coefficient $c_{0}$ is constant in time. Now we can eliminate
$c_{k}$ by integrating Eq.(\ref{four})with initial condition
$c_{k}(0)=0$ and substituting the result into Eq.(\ref{three}), then
we could obtain
\begin{equation}\label{five}
  \dot{c_{1}} = -\int_{0}^{t}d\tau G(t-\tau)c_{1}(\tau)
\end{equation}
where
\begin{equation}\label{six}
    G(t-\tau) =  \sum_{k}g_{k}^{2}e^{-i(\omega_{k}-\omega_{0})(t-\tau)}
\end{equation}
is the delay Green's function of the problem\cite{john1}.
\par In the effective mass approximation and long-time limit, the Green's function (\ref{six})takes the form\cite{john1,john2}
\begin{equation}\label{eight}
    G(t-\tau)=-\alpha\frac{e^{i[\delta(t-\tau)+\pi/4]}}{\sqrt{(t-\tau)^{3}}},~\omega_{c}(t-\tau)\gg1,
\end{equation}
under the anisotropic photon-dispersion relation\cite{john1,john2}
\begin{equation}\label{seven}
    \omega_{\mathbf{k}}\approx\omega_{c}+A(\mathbf{k}-\mathbf{k}_{0})^{2}
\end{equation}
where $\omega_{c}$ is the upper band-edge frequency and $k_{0}$ is a
constant characteristic of the dielectric material.
$\delta=\omega_{0}-\omega_{c}$ is the detuning of the atomic
frequency with respect to the band-edge frequency and $\alpha\approx
\omega_{0}^{2}d^{2}/(8 \omega_{c}\epsilon_{0}(\pi A)^{3/2})$ is a
constant that depends on the nature of the band-edge singularity,
here $d$ is the atomic dipole moment, $\varepsilon_{0}$ is the
dielectric constant in vacuum. By Laplace transform\cite{john1}, we
could finally obtain
\begin{eqnarray}
  c_{1}(t)&=& c_{1}(0)c(t)  \\
  c(t) &=& \varepsilon [ \lambda_{+}e^{i\lambda_{+}^{2}t}(1+\Phi(\lambda_{+}e^{i\pi/4}\sqrt{t}))- \\
  & & \lambda_{-}e^{i\lambda_{-}^{2}t}(1+\Phi(\lambda_{-}e^{i\pi/4}\sqrt{t}))]
  \end{eqnarray}

where

\begin{eqnarray*}
 \varepsilon&=&\frac{e^{i\delta t}}{\sqrt{\alpha^{2}-4\delta}},\\
\lambda_{+}&=&\left(-\alpha+\sqrt{\alpha^{2}-4\delta}\right)/2, \\
 \lambda_{-}&=&\left(-\alpha-\sqrt{\alpha^{2}-4\delta}\right)/2,
\end{eqnarray*}
and $\Phi(x)$ is the error function\cite{tab}. So, we could obtain
the density matrix of the subsystem as\cite{breu}
\begin{equation}
\rho(t)=  \left(
\begin{array}{cc}
  \rho_{ee}(0)|c(t)|^{2} & \rho_{eg}(0)c(t) \\
  \rho_{ge}(0)c^{*}(t) & \rho_{gg}(0)+\rho_{ee}(0)(1-|c(t)|^{2})
\end{array}
\right)
\end{equation}
\par In order to investigate the entanglement dynamics of the bipartite
system, we use Wootters concurrence\cite{woott}. For simplicity, we
set the two subsystems have the same parameters. Using the method in
Ref.\cite{bel}, we could obtain the concurrence of the whole system
as
\begin{eqnarray}
  C_{\Phi} &=& max\{0,2\sqrt{1-\beta^{2}}|c(t)|^{2}\beta\}
  \\
  C_{\Psi} &=& max\{0,2\sqrt{1-\beta^{2}}|c(t)|^{2}[\beta-\nonumber\\
  & & \sqrt{1-\beta^{2}}(1-|c(t)|^{2})]
\end{eqnarray}
when the initial states are
\begin{eqnarray}
  |\Phi\rangle &=& \beta|ge\rangle+\gamma|eg\rangle, \\
  |\Psi\rangle &=& \beta|gg\rangle+\gamma|ee\rangle,
\end{eqnarray}
respectively. Where $\beta$ is real, $\gamma=|\gamma|e^{i\varphi}$
and $\beta^{2}+|\gamma|^{2}=1$. Next, we will focus on the time
behavior of the concurrence $C_{\Phi}$ and $C_{\Psi}$ as a function
of $\beta^{2}$ and the dimensionless
quantity$\alpha^{2}t$~\cite{john1,john2}.

\begin{figure}
  \includegraphics{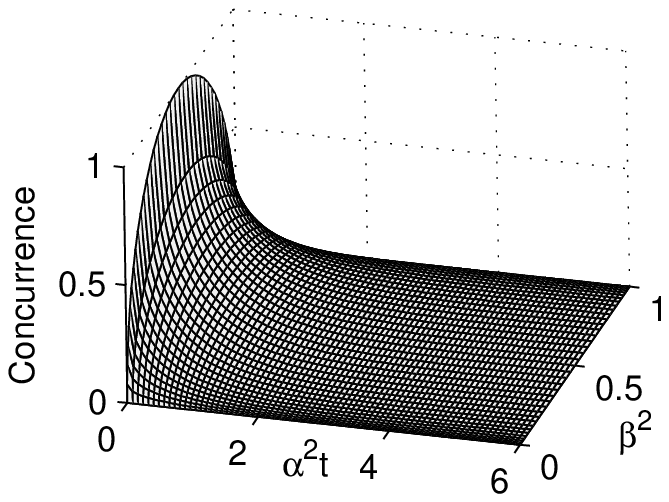}\\
  \caption{Concurrence for initial state $\Phi$ as a function of $\alpha^{2}t$ and $\beta^{2}$ in photonic crystal with $\delta=\alpha^{2}$.}\label{F1}
\end{figure}
\par First, we investigate the decoherence of two qubits initially
entangled in state $|\Phi\rangle$. Fig.~\ref{F1} shows that, with
relatively large detuning of atom frequency outside the band gap
with respect from the photonic band edge $\delta=\alpha^{2}$, the
concurrence $C_{\Phi}$ decreases exponentially to zero in time scale
for all the value of $\beta^{2}$ except for $\beta=0$ and $\beta=1$,
which corresponds to product states. The result is similar to that
case in vacuum.
\par
\begin{figure}
  \includegraphics{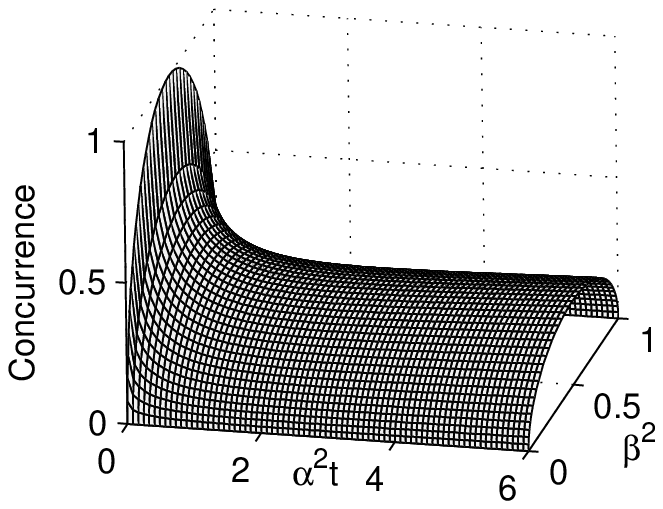}\\
  \caption{Concurrence for initial state $\Phi$ as a function of $\alpha^{2}t$ and $\beta^{2}$ in photonic crystal with $\delta=-\alpha^{2}$}\label{F2}
\end{figure}

 From Fig.~\ref{F2}, we can find that the concurrence $C_{\Phi}$, with
relatively large detuning of atom frequency inside the gap with
respect from the photonic band edge $\delta=-\alpha^{2}$, decreases
exponentially to a steady value bigger than zero for all the value
of $\beta^{2}$ except for $\beta=0$ and $\beta=1$, as time
increasing and then keep the value all the time.
\par
\begin{figure}
  \includegraphics{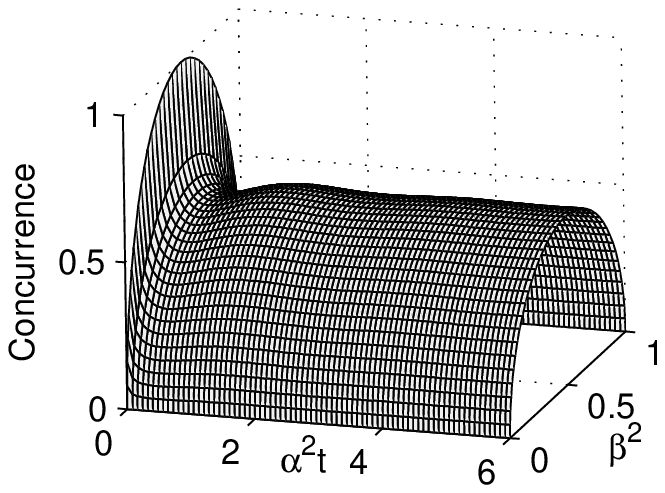}\\
  \caption{Concurrence for initial state $\Phi$ as a function of $\alpha^{2}t$ and $\beta^{2}$ in photonic crystal with $\delta=-4\alpha^{2}$.}\label{F3}
\end{figure}

 Fig.~\ref{F3} exhibits that, as the detuning of atom frequency
inside the gap with respect from the photonic band edge increases to
$\delta=-4\alpha^{2}$, the concurrence $C_{\Phi}$ will decreases to
a minimum value, and then return to a steady value. The bigger the
detuing of atom frequency inside the gap with respect from the
photonic band edge is, the bigger the steady value is.

\begin{figure}
  \includegraphics{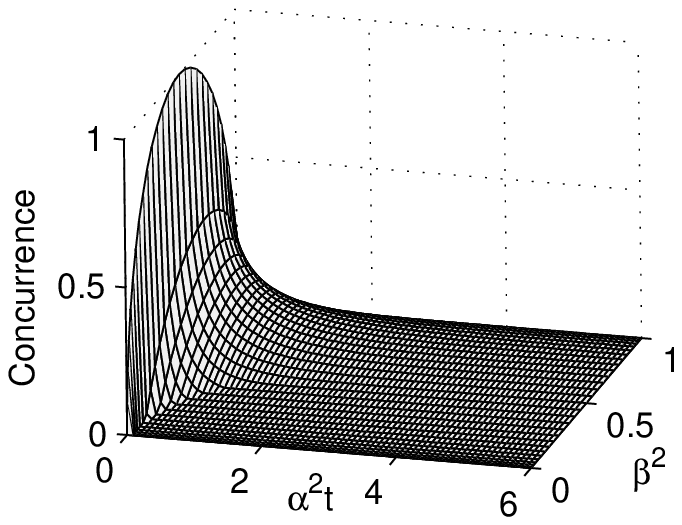}\\
  \caption{Concurrence for initial state $\Psi$ as a function of $\alpha^{2}t$ and $\beta^{2}$ in photonic crystal with $\delta=\alpha^{2}$.}\label{F4}
\end{figure}
\par Then, we investigate the decoherence of two qubits initially
entangled in state $|\Psi\rangle$. Fig.~\ref{F4} shows that, when
$\delta=\alpha^{2}$, the entanglement of the system will disappear
at finite time as $0<\beta^{2}< 0.5$ because of the dominant
influence of double excitation component $|ee\rangle$ in state
$|\Psi\rangle$, while the disentanglement occurs asymptotically in
time for the other value of $\beta^{2}$, which is similar to the
results of that case in Markovian or usual non-Markovian
regime.\cite{yu1,yu2,mf,bel}.

\begin{figure}
  \includegraphics{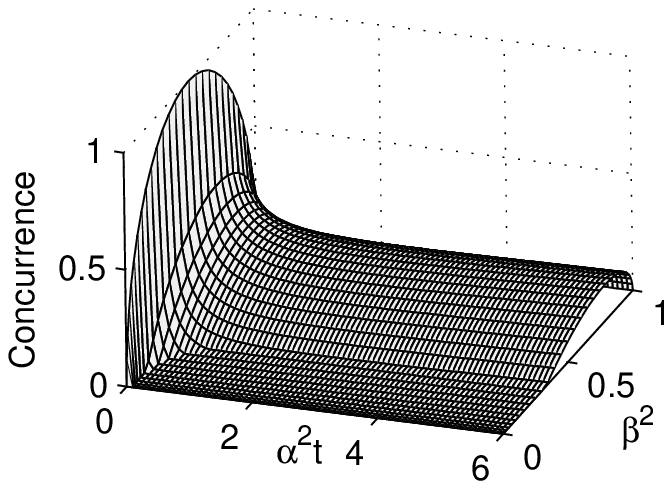}\\
  \caption{Concurrence for initial state $\Psi$ as a function of $\alpha^{2}t$ and $\beta^{2}$ in photonic crystal with $\delta=-\alpha^{2}$.}\label{F5}
\end{figure}
\par Fig.~\ref{F5} reveals that, for $\delta=-\alpha^{2}$, the entanglement of the system will disappear
at finite time as $0<\beta^{2}\leq 0.35$, while the entanglement for
the other value of $\beta^{2}$ will decreases exponentially to a
steady value bigger than zero as time increasing, which is similar
to the cease of $C_{\Phi}$ in Fig.~\ref{F2} for the same parameter.
\begin{figure}
  \includegraphics{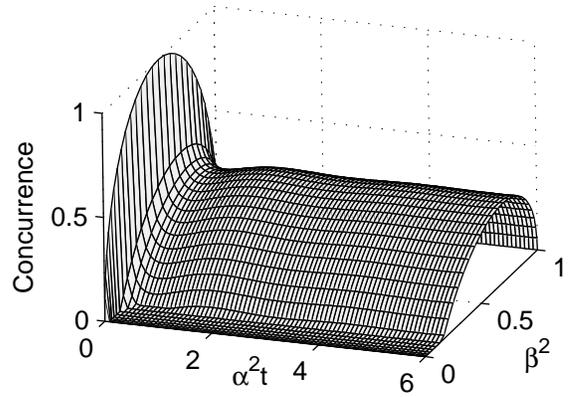}\\
  \caption{Concurrence for initial state $\Psi$ as a function of $\alpha^{2}t$ and $\beta^{2}$ in photonic crystal with $\delta=-4\alpha^{2}$.}\label{F6}
\end{figure}

\par From Fig.~\ref{F6}, it is evident that the range of
$\beta^{2}$, for finite time disentanglement, becomes obviously
smaller than that for the cease in Fig.~\ref{F5}. And the time
evolution behavior of concurrence $C_{\Psi}$, in the other range of
$\beta^{2}$, is similar to the cease of $C_{\Phi}$ in Fig.~\ref{F3}
for the same parameter. The bigger the detuing of atom frequency
inside the gap with respect from the photonic band edge is, the
smaller the finite time disentanglement range of $\beta^{2}$ is.
\par The above results show that the time behavior of decoherence,
as $\delta<0$, is greatly different from that of the ceases in
Markovian and usual non-Markovian regime\cite{yu1,yu2,mf,cao,bel}
because of the existence of photonic band gap.

\begin{figure*}
  \includegraphics{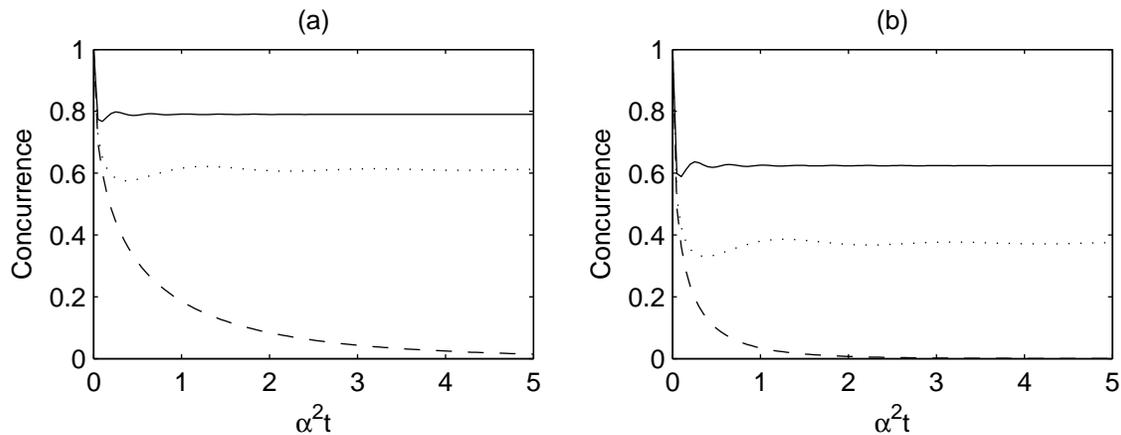}\\
  \caption{Concurrence for initial states (a) $(|ge\rangle+|eg\rangle)/\sqrt{2}$ and (b) $(|gg\rangle+|ee\rangle)/\sqrt{2}$. solid line:
$\delta=-20\alpha^{2}$ dotted line:  $\delta=-5\alpha^{2}$ dashed
line: $\delta=0.5\alpha^{2}$.}\label{F7}
\end{figure*}

\par Last, we will compare the time behavior of the concurrence of maximum entanglement
state $(|ge\rangle+|eg\rangle)/\sqrt{2}$ with that of
$(|gg\rangle+|ee\rangle)/\sqrt{2}$. From Fig.~\ref{F7}, it is
obviously that decoherence behavior for the case of initial state
$(|ge\rangle+|eg\rangle)/\sqrt{2}$ is similar to that of
 $(|gg\rangle+|ee\rangle)/\sqrt{2}$ under the same parameter. As the
 atomic frequency is detuned inside the gap $\delta<0$, the concurrence for (a)
 and (b) in Fig.~\ref{F7} will decrease to steady value respectively, besides the steady value for the case of initial state
$(|ge\rangle+|eg\rangle)/\sqrt{2}$ is bigger than that of
 $(|gg\rangle+|ee\rangle)/\sqrt{2}$ under the same parameter. And the decay rate of concurrence for the case of initial state
 $(|gg\rangle+|ee\rangle)/\sqrt{2}$, as $\delta\geq0$ , is bigger than that for the case of initial state
$(|ge\rangle+|eg\rangle)/\sqrt{2}$. So we could conclude that the
state $(|ge\rangle+|eg\rangle)/\sqrt{2}$ is more robust than the
state $(|gg\rangle+|ee\rangle)/\sqrt{2}$ against quantum noise in
vacuum.
\par In summary, we have analytically derived the concurrence of two initially entangled atoms coupled individually to its own anisotropic band gap photonic
crystal environment without Born or Markovian approximation. The
results show that the finite time decoherence invoked by spontaneous
emission could be suppressed efficiently for the relatively large
detuning of atom frequency inside the gap with respect from the
photonic band edage, and the entanglement of Bell state
$\beta|ge\rangle+\gamma|eg\rangle$ is more easily preserved in
photonic crystal than that of $\beta|gg\rangle+\gamma|ee\rangle$
under the same condition.
\par In practice, we could store many
particle pairs of maximum Bell entanglement state as
$(|ge\rangle+|eg\rangle)/\sqrt{2}$ in photonic crystal individually.
After an interval of time, all the particle pairs become partial
entanglement particle pairs. However, we could recover the maximum
entanglement by concentrating the partial entanglement with local
operation\cite{ben}. Then, we could get entangled particle pair in
any of the four maximum Bell entanglement states by local operation
on particle pair in state of
$(|ge\rangle+|eg\rangle)/\sqrt{2}$~\cite{ben1}.

\begin{acknowledgments}
This work was supported by the National Natural Science Foundation
of China Grants No.60578055, the State Key Program for Basic
Research of China under Grant No. 2007CB925204 and No.2007CB307001.
\end{acknowledgments}

\newpage 
\bibliography{apssamp}

\end{document}